\documentclass[a4paper,11pt]{article}
\usepackage{pos}

\newcommand{\apj}{ApJ}

\newcommand{\aap}{A\&A}

\newcommand{\aj}{AJ}
\newcommand{\mnras}{MNRAS}

\newcommand{\HI}{H{\sc i}}

\newcommand{\sunn}{$_{\odot}$}

\title{Nearby Voids and Their Galaxies: Recent Progress and~Prospects}
\ShortTitle{Nearby Voids and Their Galaxies}

\author*[a]{S.~Pustilnik}
\author[a]{Y.~Perepelitsyna}
\author[a]{A.~Tepliakova}
\author[a,b]{A.~Kniazev}
\author[a,b]{E.~Egorova}
\author[c]{J.~Chengalur}
\author[d]{S.~Kurapati}

\affiliation[a]{SAO RAS, Nizhny Arkhyz, Russia}
\affiliation[b]{SAI, Moscow State University, Universitetsky Pr. 13, Moscow, Russia}

\affiliation[c]{NCRA, Ganeshkind, Pune, India}

\affiliation[d]{Department of Astronomy, University of Cape Town, Rondenbosch 7701, South Africa}

\emailAdd{sap@sao.ru}
\abstract{
Voids occupy about 3/4 of the volume of the Universe
and contain about 15\% of its mass. Due to various
observational selection effects, these structural elements
and galaxies populating the voids, are highly under-explored.
This especially  relates to the lowest mass galaxies which
comprise the main void population. Studying the nearby voids
allows us to improve our understanding of the most elusive
void objects.
We present a brief overview of the current status and
prospects of the study of nearest voids and their galaxies.
First, we summarise the pioneer study of a hundred galaxies
residing in the nearby Lynx--Cancer void, which clearly evidences
for the slower evolution of void galaxies and also finds unusual
very metal-poor and gas-rich dwarfs. Then we describe the
recently defined sample of nearby voids within a
sphere with R = 25 Mpc and a sample of 1350~galaxies residing
in these voids ($\sim$20\% of all the galaxies within this volume). We
discuss the current results obtained for several directions of
the study  of this sample. They include: the search for Very
Young Galaxies, the study of HI properties,  the clustering of void
galaxies and its relation to the void substructures, and the
unbiased study of 260 void galaxies within the Local Volume
(R $<$ 11 Mpc). Altogether, this opens a promising way to address
the suggested peculiarities of void galaxy formation and evolution. Finally,
we briefly overview the expected advancements in the void
galaxy studies related to the upcoming  new facilities.
}

\FullConference{%
  The Multifaceted Universe: Theory and Observations -  2022 (MUTO2022)\\
  23-27 May 2022\\
  SAO RAS, Nizhny Arkhyz, Russia\\}

\begin{document}
\maketitle

\section{Introduction. Voids and their galaxies}

Voids represent one of the four types of cosmic web elements,
ranked according to the decreasing matter density: nodes, filaments, walls, and voids (e.g.,
Cautun et al., 2014 \cite{Cautun2014}). Having a matter density of $\sim$1/5 of the mean density
of the Universe,
voids are unique objects for cosmology since they
have substructure still growing in the linear regime so that they can be
treated as ``time machines'' and ``cosmological microscopes'' (Aragon-Calvo \&
Szalay, 2013 \cite{Aragon13}).

The Nearby Void Galaxy Sample (\mbox{d $<$ 25} Mpc) (Pustilnik et al., 2019 \cite{PTM19})
and its subsample
within the Local Volume  (\mbox{d $<$ 11} Mpc) directly relate to the near-field
cosmology. They allow us to probe in the greatest detail the processes of the Mpc-scale
structure  formation from the initial perturbations and to witness the recent
galaxy build-up.

\section{Three main directions in void studies}

1. Statistics of observed voids as an ensemble of separate entities.
Comparison to cosmological models and simulations.

2. Studying of void substructure and dynamics. This is based on model simulations.
To address these issues observationally, it takes to identify
many galaxies per void as tracers of the substructure. Currently,
this is beyond the available opportunities. However,
the expected technological advancements, related to the upcoming
new instruments (mentioned later), will open this observational direction.

3. Formation and evolution of void galaxies.  Currently, we already have 
reasonable observational support of these studies with large telescopes.
However,  the future
new facilities will greatly advance this direction as well. This will allow us,
in particular, to address the issue of ``dark'' protogalaxies and the minimal
baryonic mass of galaxies survived cosmic reionization.

\section{Previous studies of voids and their galaxies}

To study void structures and galaxy evolution,  we need large and deep
samples  of  void galaxies.
Previous mass studies of void galaxies dealt mainly with the large ``distant'' voids
(\mbox{$D\sim100$--20~Mpc}),  in which
the SDSS-based galaxy samples probed the top of the void luminosity function
(\mbox{$-20 < M_{\rm B,r} < -17$}) (e.g. Rojas et al. 2005 \cite{Rojas05}, Kreckel et al.
 2012 \cite{Kreckel12}).
The caveats of these works are the shallowness in galaxy luminosities
and the small number of galaxies per void.
Differences in the properties of void and wall galaxies for the brighter
void galaxies are small: void objects are more gas-rich, and have higher SFR.
This suggests that the ``massive'' part of  void galaxies are less sensitive to
the global environment.

Density of galaxies grows from the void centre to its border.
The central density is $\sim$0.1 of the mean density in the Universe.

The main (dwarf, say M$_{\rm B} > -16$) void population remained till the
recent time largely unexplored. However, our earlier study
evidences for the slower average evolution and on the unusual properties
of a part of void objects (e.g. Pustilnik et al. 2010, 2011,
2016 \cite{J0926, PaperIII, PaperVII}).

\section{Importance of the studying of nearby voids}

To study void structure, we need many galaxies which delineate it.
Due to the raising galaxy luminosity function, the number of galaxies grows with
decreasing luminosity. The effect of environment also gets stronger for lower
mass galaxies. Therefore, to study substructure of voids, and to understand the diversity
of their galaxy properties and evolutionary scenarios, we need large and deep galaxy samples.

In typical wide-field redshift surveys, the limiting apparent magnitudes of B$_{\rm tot} \sim 18-19$
allows one to collect faint void dwarfs (to M$_{\rm B} = -10$, $-12$ mag)
only in the surroundings of the Local Volume (LV), at distances of $\lesssim$ 20--25 Mpc.
Thus, studying of low-mass dwarfs in voids dictates the need of defining void regions
adjacent the LV.

\subsection{Brief summary of the study of the Lynx--Cancer void galaxy sample}

A sample of a hundred galaxies in the nearby Lynx--Cancer void (d$_{\rm centre} \sim$ 18~Mpc)
was formed (Pustilnik \& Tepliakova 2011 \cite{PaperI}) and studied in a series of 10 papers
 (Pustilnik et al. 2016 \cite{PaperVII} and references
therein). The main results and conclusions of this study are as follows:
(a) the major fraction of void galaxies are low surface brightness (LSB) dwarfs;
(b) the average M(gas)/L$_{\rm B}$ is a factor of $\sim$1.4 larger than that of the reference sample;
(c) gas metallicity Z(gas) at a fixed M$_{\rm B}$ is reduced, on average, by a factor of $\sim$1.4 (in comparison
to the LV reference sample from Berg et al., 2012 \cite{Berg12});
(d) in addition, there exists a small group of ``unusual'' low-mass dwarfs.
They have extremely low Z(gas) $\sim$Z\sunn/50--Z\sunn/30, reduced by a factor of 2--5 for their
luminosity, very high M(gas)/M(bary) $\sim$0.97--0.99, and blue colours of the periphery,
corresponding to the ages of the oldest visible stellar population of 1--3 Gyr.

{\bf Conclusions.}
(a) Void galaxies as a whole, on average appear less evolved. This implies either the
slower secular evolution  or/and delayed galaxy formation.
(b) There exists a small number ($\sim$10\%) of the least massive void dwarfs with
very low gas metallicities and
other extreme properties, indicating their early stages of evolution.

\section{eXtremely Metal-Poor (XMP) galaxies. Relation to voids}

XMP galaxies (conditionally with Z(gas) $<$ Z\sunn/30) are very rare in the nearby Universe.
Since the discovery in 1970 of the first such object, the blue compact galaxy IZw18
(Sargent \& Searle, 1970 \cite{IZw18}), there have been numerous attempts to find more
such galaxies. The main motivation was
that they  present a unique opportunity to study in detail the processes
in metal-poor gas and stars typical of the conditions in galaxies in the early Universe.
Only a handful of such galaxies were identified among hundreds of thousands
star-forming emission-line galaxies, mainly from the SDSS spectral survey
(e.g., Izotov et al. 2019 \cite{Izotov2019}).

As the follow-up analysis of the spatial positions of the known XMP dwarfs has shown,
the majority of them, including the prototype IZw18, reside in voids. Having in mind also our
results on the galaxy population in the Lynx-Cancer void, it was clear that the search
for XMP galaxies in voids was promising.

\section{Advancement in the studying of nearby voids and their galaxies}

Based on the above results, it was tempting to extend the sample of a hundred
galaxies in the Lynx-Cancer void
and to study the phenomenon of void galaxies on a much greater statistical ground.

1. In 2017, we identified 25 voids within a sphere with R $<$ 25 Mpc and
  formed a sample of 1350 galaxies residing in these voids (Nearby Void Galaxies, NVG,
Pustilnik et al. 2019 \cite{PTM19}).

2. In 2017--2020, we undertook a search for new XMP dwarfs among the least
luminous NVG objects, namely those with M$_{\rm B} > -14.2$~mag. From 60 additionally preselected
candidates, we found via BTA and SALT spectroscopy 30 new dwarfs with
Z(gas) = (0.02--0.04) Z\sunn\  (Pustilnik et al. 2020, 2021 \cite{XMP-SALT, XMP-BTA}).
These dwarfs serve as a basement for studying the diversity of low-metallicity galaxies and
various issues of galaxy evolution in voids.  Part of them resemble in their properties
predicted in simulations the so-called Very Young Galaxies (VYG, Tweed et al. 2018 \cite{Tweed2018}),
in which the majority of stars are formed during the last $\sim$1~Gyr.

\section{New project: studying void galaxies in the Local Volume}

The Local Volume (LV, R $<$ 11~Mpc) and the sample of the LV $\sim$1250 galaxies
(Karachentsev et al. 2013 \cite{UNGC}) represent
a very important reference sample used for comparison with models of the nearby Universe
in numerical cosmological simulations. However, there is a significant biasLow-redshift lowest-metallicity star-forming galaxies in the SDSS DR14
in the observational studies of the sample galaxies, which mostly include those from typical groups.
This bias was caused by the absence of the information on voids in the LV, with except of the well known
giant Tully void with $\sim$15 galaxies catalogued within its boundaries.

The creation of the nearby void galaxies sample \cite{PTM19}
allowed us to trace the part of the nearby voids filling the Local Volume.
As a result, we separated a subsample of 260 void galaxies residing within the LV.
The study of this sample is being conducted in several stages.

The first stage (2020--2022), which is almost finished, started
from a comparative study of gas metallicity in
61 the least luminous (M$_{\rm B} > -13$~mag) void galaxies within the LV with that
for the reference sample of the late-type galaxies in the LV from \cite{Berg12}.
The second, ongoing, stage (2022--2024?) includes the spectral study of all the 130 LV void
galaxies with M$_{\rm B} > -14.3$~mag, which is the median of the
luminosity distribution for this sample.

For the great majority of the studied void galaxies, their Z(gas) falls into the
low-metallicity region (conditionally, Z(gas) $<$ Z\sunn/5, or 12+log(O/H) $<$ 8.0 dex).
However, only for a minor part of the studied void galaxies, O/H(gas) was determined via the
direct method when the temperature-sensitive faint line [O{\sc iii}]$\lambda$4363 was detected.
For the remaining objects, O/H was determined via empirical estimators based on the strong lines
of Oxygen and Hydrogen.

{\bf Intermediate summary}

To date, spectra of $\sim$76 galaxies of this subsample are available, mostly thanks
to our observations. Their gas metallicity appears to be in the range of 12+log(O/H) = 7.05--8.0 dex.
Among others, several new XMP dwarfs are discovered.

\begin{figure*}
\centering{
\includegraphics[width=12.0cm,angle=-90,clip=]{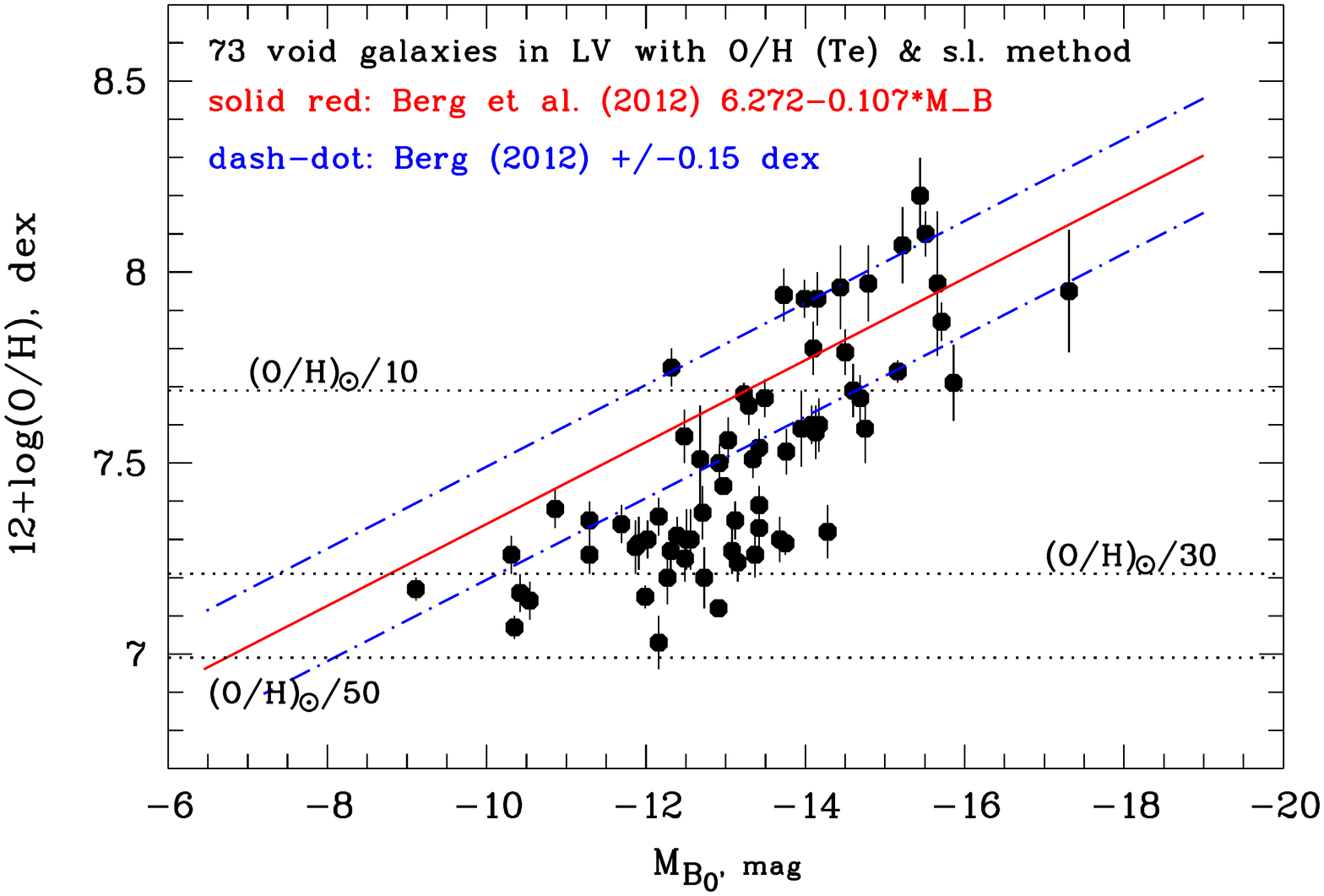}
\caption{\label{fig:ZvsMB}
Relation of 12+log(O/H) versus M$_{\rm B}$ for 73 void galaxies
residing in the LV: black octagons with error bars. The red solid line shows the linear
regression for the reference sample from the LV sample \cite{Berg12}.
Two parallel dash-dotted lines show the r.m.s. scatter of the reference sample
around the linear regression ($\pm$0.15~dex).
} }
\end{figure*}

In Figure~\ref{fig:ZvsMB} we show the relation of 12+log(O/H) versus M$_{\rm B}$
for 73 LV void galaxies with the most reliable data in
comparison with the reference sample of late-type galaxies from Berg et al. 2012 \cite{Berg12}.
It is drawn with the solid line along with two parallel dash-dotted lines illustrating
the scatter of the log(O/H) in the reference sample around the linear regression ($\pm$0.15~dex).
As discussed in Pustilnik et al. (2016, 2021) \cite{PaperVII, XMP-BTA},
void galaxies have, on average, a reduced
value of gas O/H. Furthermore, its scatter is substantially elevated.

\section{Large scatter of metallicity in void galaxies. Clustering in
voids and its possible implications}

The elevated scatter of gas metallicity for a given M$_{\rm B}$ in void galaxies
seemingly indicates additional factors affecting secular evolution.
It is also possible that for a fraction of the observed void galaxies their reduced
gas metallicity can be a short-term effect due to the localized induced star formation
episodes related to the accretion of a ``primordial'' gas blob (e.g. Ceverino et al., 2016
\cite{Ceverino16}).
The galaxies of the reference sample from Berg et al. (2012) \cite{Berg12} belong mostly to the
typical groups of the LV. Their mutual interactions induce elevated
star formation and, thus, lead to additional production of metals.

Void galaxies cluster similar to non-void ones, but with 
substantially  reduced amplitudes on all scales. Depending on their
local environment, their properties can vary significantly.
That is the observed properties of void galaxies are the product
of the interplay between the global and local environments.

There are several topics for a deeper insight on the void galaxy clustering
and the related peculiarities of their evolution.  They include the following:
the identification of various galaxy associations (pairs, triplets,
quartets, and other aggregates, including a dominant host galaxy)
as the probable nodes of the void web;
the identification of the unbound galaxies at mutual
distances of $<$ 0.5--1~Mpc as the probable tracers of void filaments.

Chemical evolution in the ``massive'' aggregates in voids may be
more similar to that in the non-void groups.
E.g., sub-luminous hosts in voids can trigger SF and
accelerate their companion's evolution via tidal interaction.

There are hints on the existence of XMP dwarf associations in voids.
That is some void regions can appear the birthplaces of ``young'' galaxies.
This would be an important finding. However, a much larger statistics
of such XMP dwarf associations is needed to get reliable results.

\section{Gas in void galaxies}

\begin{figure*}
\centering{
\includegraphics[width=4.7cm,angle=-0,clip=]{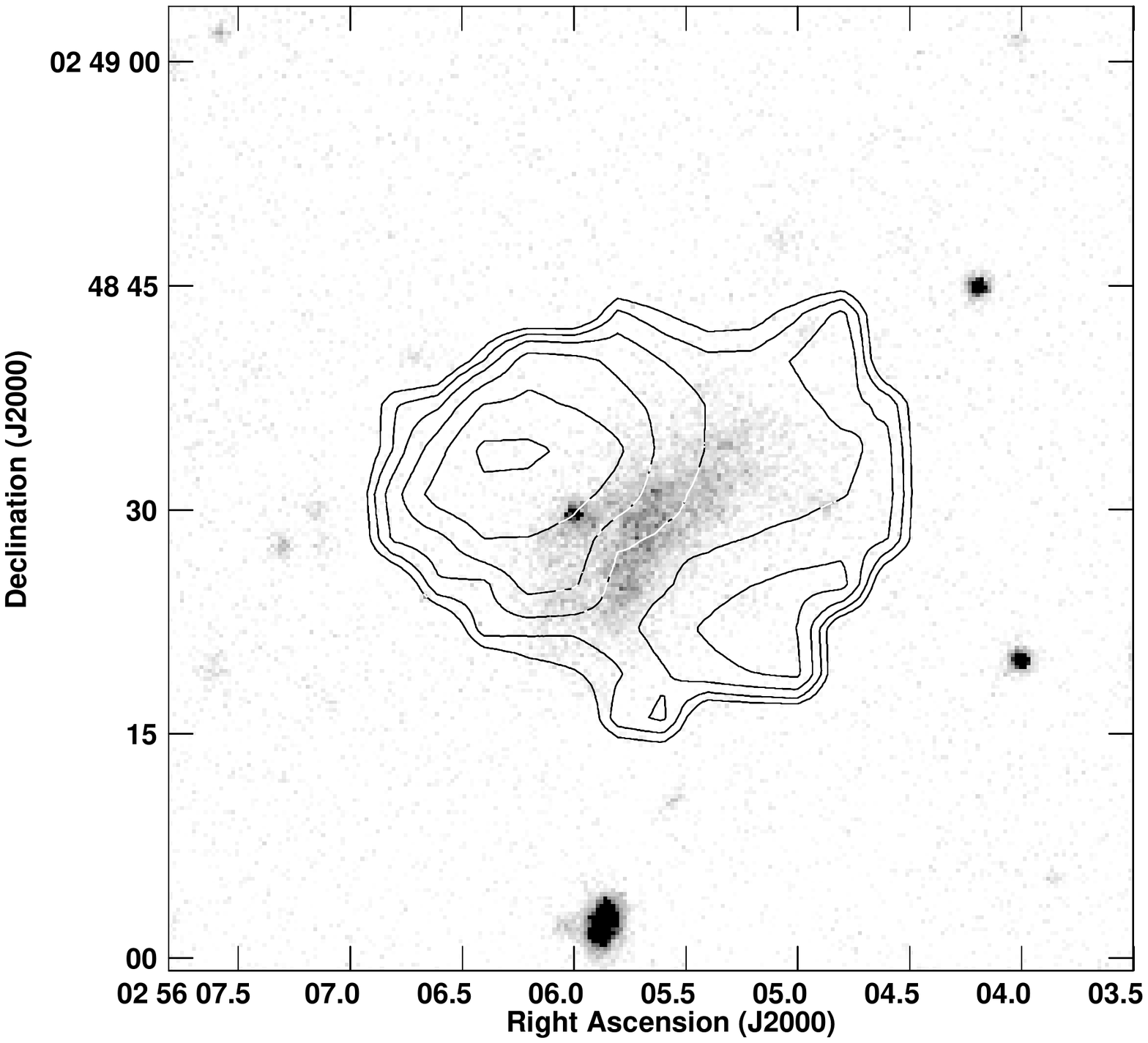}
\includegraphics[width=4.7cm,angle=-0,clip=]{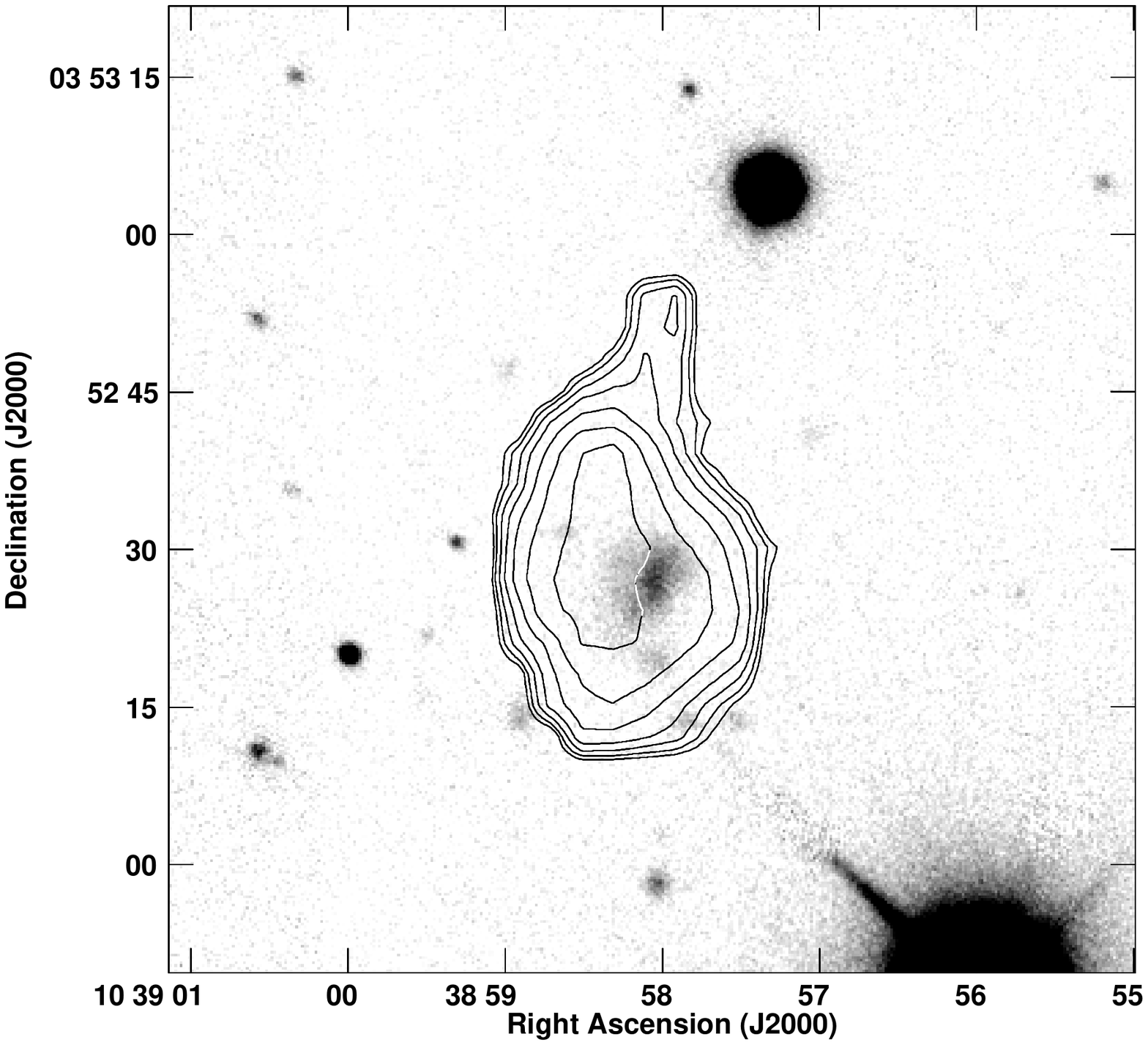}
\includegraphics[width=4.7cm,angle=-0,clip=]{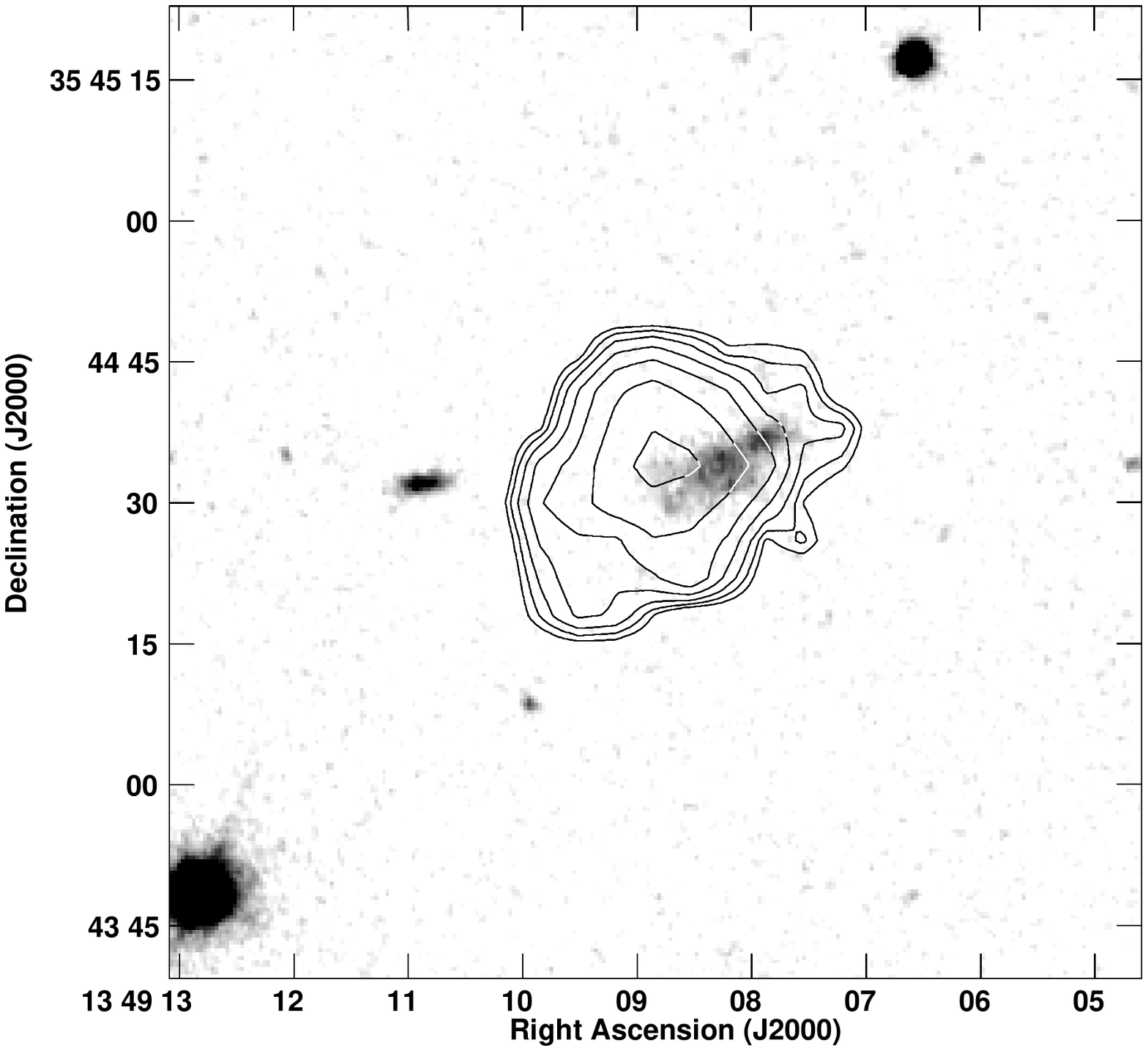}
\caption{\label{fig:XMP-HI-GMRT}
The GMRT \HI-maps with an angular resolution of $\sim$10~arcsec overlaid on
the SDSS images of three isolated XMP galaxies in the nearby voids.
From left to right: the galaxies AGC124629, AGC208397, and AGC239144.
} }
\end{figure*}

Atomic gas (\HI\ and HeI) in the majority of void galaxies appears to be the
dominant component of their baryonic matter. Therefore, its study is
an obligatory step in order to understand the properties of void objects.
We mapped tens gas-rich void dwarfs in the \HI 21-cm line using the Giant Metrewave Radio Telescope
(GMRT).
In Figure~\ref{fig:XMP-HI-GMRT} we present such maps for three isolated XMP dwarfs
in the nearby voids. The disturbed morphology of \HI\ gas in AGC124629,
AGC208397, and AGC239144 indicates its
non-equilibrium state and hints the continuing build-up of their gas body.
This observed phenomenon seems to get support from modern
high-resolution cosmological simulations of void galaxies.
At the same time, this unprocessed ``primordial'' inflowing gas should
reduce the total galaxy metallicity,
playing as a factor of slower secular evolution.

\section{Study of nearby voids: new horizons}

The new facilities coming during the next 5--10 years will
qualitatively advance studies in the area.

1. The SKA and ngVLA will increase the sensitivity to the low-baryon-mass
objects by orders of magnitude, increasing the current $\sim$1300 known
Local Volume galaxies, 260 of them in voids, to tens of thousands LV objects and
thousands of void galaxies with masses
M(HI) $\lesssim 10^{6}$ M\sunn\ (if they exist).
This progress will make it possible: \\
-- to probe the mass function to this limit, \\
-- to uncover the tenuous substructure of voids, \\
-- to discover ``dark'' galaxies (conditionally with M*/M(gas) $<$ 0.01), \\
-- to address SF in very atypical conditions. \\

2. The JWST and then the E-ELT and TMT will expand the tip-RGB distances to at
 least D$\sim$25 Mpc, greatly improving the accuracy of void galaxy positions,
 and will allow us to probe the detailed SF histories and
fractions of the old and younger stellar populations.

{\bf New deep optical and NIR surveys:}

3. The DES imaging survey and its clones and deep NIR surveys will
allow us to better determine the parameters of global stellar populations via
multi-colour photometry and SED fitting.

4. Extending of the SDSS spectral survey to the Southern
hemisphere will substantially improve spectral data and
open the opportunity to address the metallicity of many new void
dwarfs and improve the completeness of the census of nearby voids.

\section{Conclusion}

The deep studying of dwarf galaxies in the nearby voids
is in many aspects challenging even for the largest ground-based telescopes.
However, thanks to their proximity, this gives us valuable information
in intimate details of galaxy evolution and star formation at the extreme
conditions partly related to the conditions in the early Universe.

{\bf Acknowledgements} \\
The work on the presented research programs has been supported during the recent years by grants
of the Russian Foundation for Basic Research (No.~18-52-45008) and the Russian Science
Foundation (No.~22-22-00654, the part related to the Local Volume void galaxies).

\end{document}